\documentclass[letterpaper,12pt, abstracton]{scrartcl}

\usepackage{marvosym}
\usepackage[T1]{fontenc}
\usepackage{color,hyperref}
\usepackage{enumitem}
\usepackage{amssymb}
\usepackage{amsmath}
\usepackage{wasysym}
\usepackage{pifont}
\usepackage{makecell}
\usepackage{subcaption}
\usepackage{booktabs}
\usepackage{Sweave}

\newcommand\Rey{\mbox{\textit{Re}}}  
\newcommand\Wi{\mbox{\textit{Wi}}}  
\date{}

\begin{document}

\title{Is a UCM fluid flow near a stationary point always singular? - Part II}
\author{Igor Mackarov \\ \href{mailto:Mackarov@gmx.net}{{\small Mackarov@gmx.net}}}

\maketitle
\begin{abstract}
Oftentimes observed divergence of numerical solutions to benchmark flows of the UCM viscoelastic fluid is a known and widely discussed issue. Some authors consider such singularities ``invincible''. Following the previous research, the article gives more arguments against this position, for which it considers two typical flows with a  stagnation point, often a place of the flows' singularity. For the flow spread over a wall, as previously for the counterflows, numerical and asymptotic analytical solutions are presented. Both kinds of flows turn out regular in the stagnation points, in particular, for high Weissenberg numbers. A good accordance is demonstrated between the analytical and numerical results.


\end{abstract}

\section{Introduction}
\label{intro}
\label{Intro}
Benchmark flows of viscoelastic fluids such as a flow near a flat wall or past a cylinder or counterflows within cross slots often provide a number of  features difficult to describe \cite{Nonlinearity}, such as
\begin{itemize}[label=\textbullet]
	\item formation of various kinds of secondary flows or vortexes often leading to purely elastic instabilities \cite{Benchmark,Groisman97,Groisman98,Pathak,LiXi},
	\item spontaneous rise of flow asymmetries near stationary points \cite{Couette,MackarovBif,Mackarov2014,Poole,Rocha},
	\item  flows reversals \cite{Mackarov2011,Mackarov64,Renardy2006} (more of which will be shown here),
	\item emergence of singularities in a flow, in particular, significant	strains in the vicinity of a stagnation point yielding local zones of extremely high stresses --- sometimes in the form of boundary layers and birefringent strands \cite{Becherer,Objections,Renardy2000,Thomases,Vajravelu,Wapperom}.
\end{itemize}

As to the last item, it is common to observe appearance of  singularities in special points of an important class of \textit{stagnation flows} \cite{Pipea2009,LiXi}. With respect to the UCM model, some authors tend to consider such emergence ``near special points where the velocity vanishes --- even though the geometry is not singular'' \cite{Objections} --- an intrinsic value of this model. Regarding some flows, doubts are raised about overall possibility to satisfy the UCM rheological law and momentum equation simultaneously \cite{VanGorder}.

Here we are going to dispute the inevitability of the stresses singularity in stationary and stagnation points. For this we will briefly repeat the main conclusions from the analytical and numerical study of counterflows within cross slots \cite{Mackarov_arxiv}. Such a comparative study  will be in detail repeated for the spread along a wall. The existence of regular regimes of the spread flow near the stationary point will be demonstrated with an emphasis on the computation--analytics correlation and a close look at convergence of the numerical solutions (in particular, for high Weissenberg numbers).

\begin{figure*}[t]
	\centering
	\begin{subfigure}[]{0.45\textwidth}
		\includegraphics[width=1\textwidth]{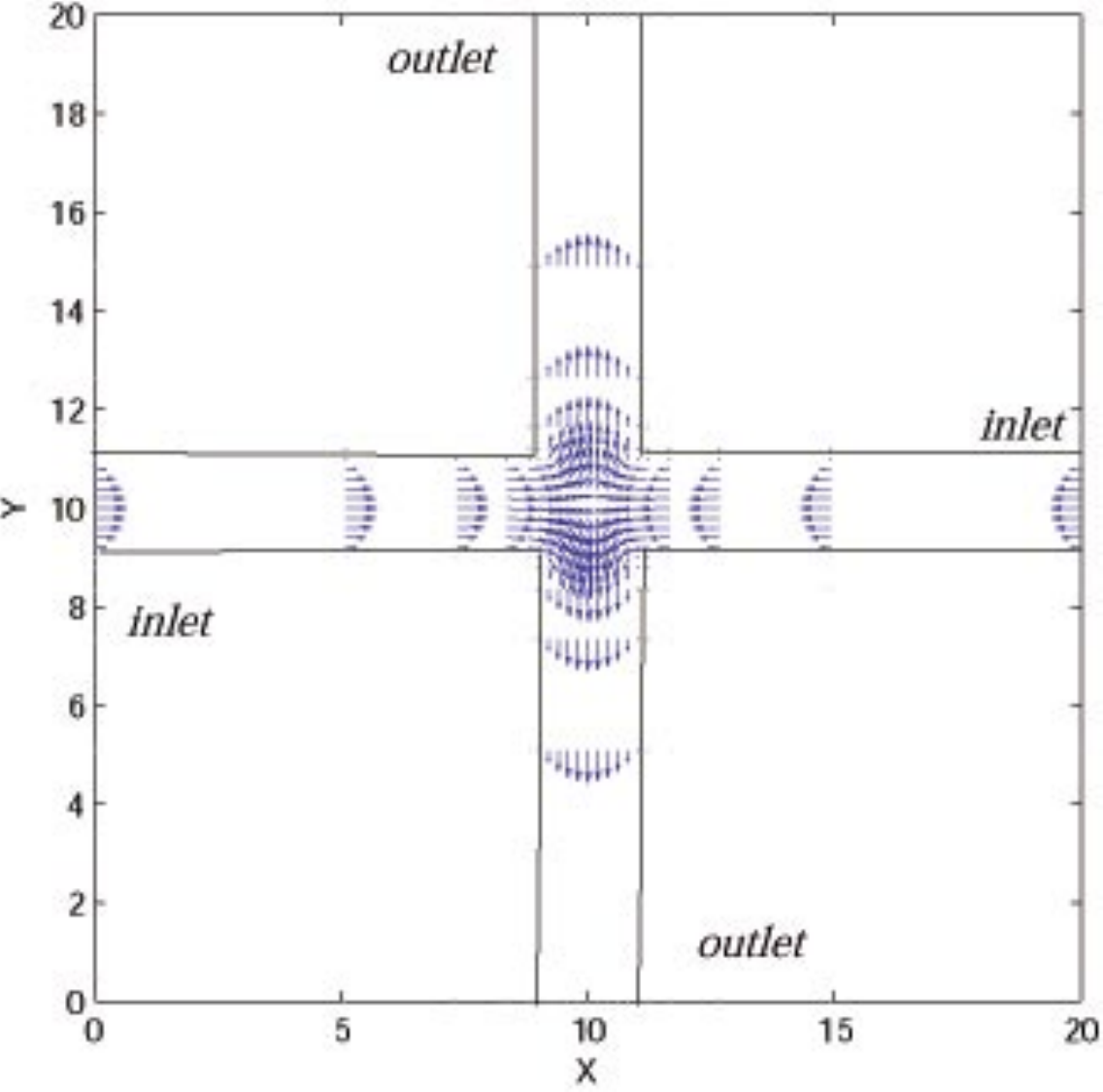}
		\caption{counterflows}
		\label{fig:counterLayout}
	\end{subfigure}%
	~ 
	\begin{subfigure}[]{0.45\textwidth}
		\includegraphics[width=1\textwidth]{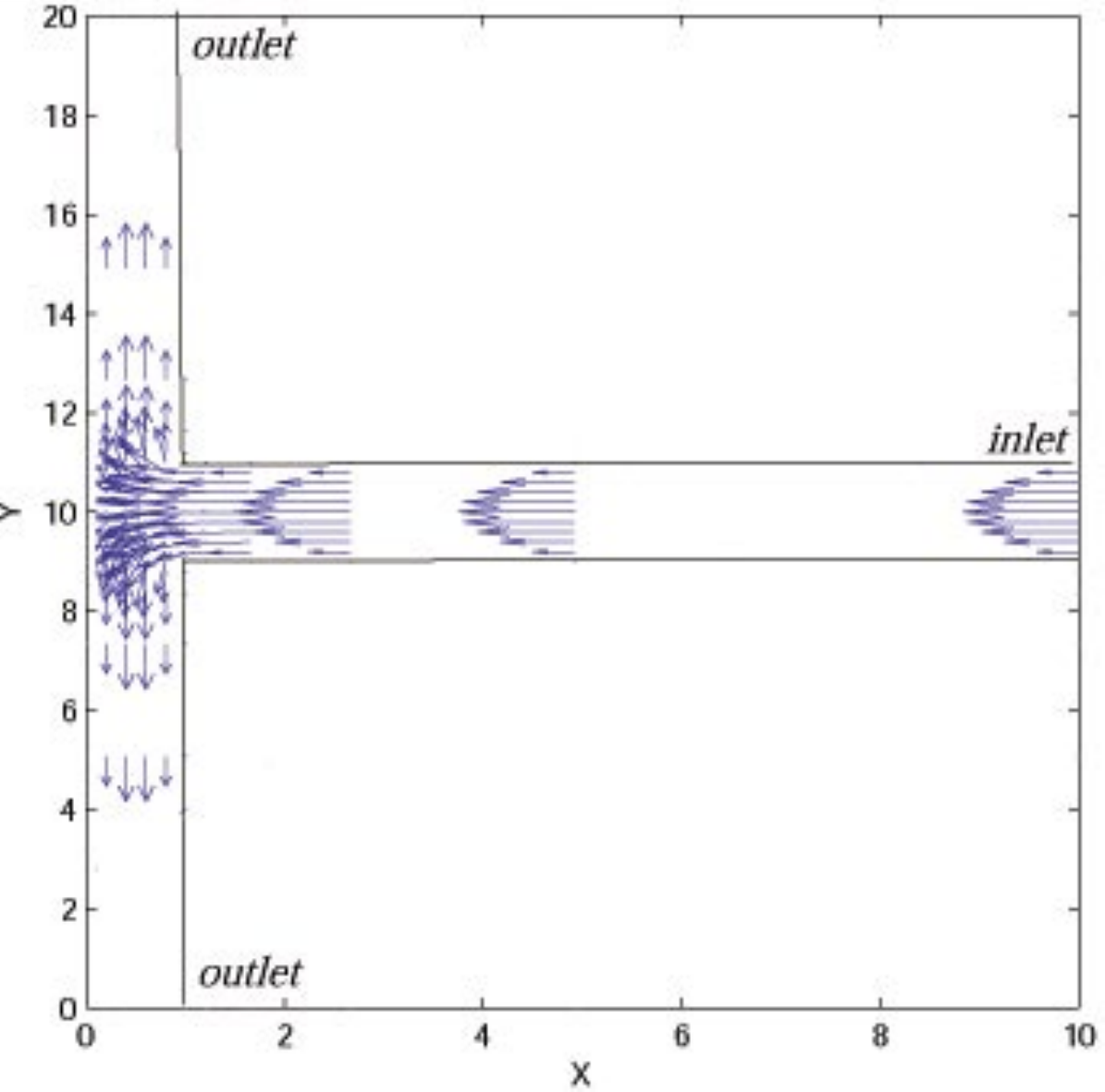}
		\caption{spread over a wall}
		\label{fig:spreadLayout}
	\end{subfigure}
	\caption{General layout of the flows considered.}
	\label{fig:layout}
\end{figure*}

\section{The problems statement}
Both problems --- of counterflows moving over two pairs of slots and of a flow propagating over a slot to spread at a wall (Fig.~\ref{fig:layout}) --- imply a common formal statement.

In terms of the dimensionless variables normalized on the problem natural scales (asymptotic stationary inlet pressure ${p_{inlet}}$, the fluid density $\rho$, the velocity scale $\sqrt {{\raise0.7ex\hbox{${{p_{inlet}}}$} \!\mathord{\left/
			{\vphantom {{{p_{inlet}}} \rho }}\right.\kern-\nulldelimiterspace}
		\!\lower0.7ex\hbox{$\rho $}}} $~, and a horizontal slot semi-width) the UCM constitutive equation can be written as

\begin{equation}
	\frac{{{D_O}\,{\sigma _{ij}}}}{{{D_O}\,t}}  + \frac{1}{{\Wi}}{\sigma _{ij}} =  \,\frac{1}{2}\frac{1}{{\Wi \Rey}}\left( {\frac{{\partial {v_i}}}{{\partial {x_j}}} + \frac{{\partial {v_j}}}{{\partial {x_i}}}} \right),\,\,\,\,\,\,\,\,\,\,\,\,i,\,j = 1,2.
	\label{UCM}
\end{equation}

\noindent Here the Oldroyd derivative of the stress tensor (henceforth summation is implied over the repeated indexes ) is
\begin{equation}
	\frac{{{D_O}\,{\sigma _{ij}}}}{{{D_O}\,t}} \equiv \frac{{\partial {\sigma _{ij}}^{}}}{{\partial \,t}} + {v_k}{\frac{{\partial \,{\sigma _{ij}}}}{{\partial \,{x_k}}}_{}} - \frac{{\partial \,{v_i}}}{{\partial \,{x_k}}}{\sigma _{kj}}^{} - \frac{{\partial \,{v_j}}}{{\partial \,{x_k}}}{\sigma _{ik}},\,\,\,\,\,\,\,\,\,\,\,\,i,\,j,\,k = 1,2.
	\label{Jauman}
\end{equation}

\noindent The flows are also governed by the equations of momentum
\begin{equation}
	\frac{{\partial \,{v_i}}}{{\partial \,{t}}} + {v_j}\;\frac{{\partial \,{v_i}}}{{\partial \,{x_j}}} =  - \frac{{\partial p}}{{\partial \,{x_i}}} + \frac{{\partial {\kern 1pt} {\sigma _{ij}}}}{{\partial \,{x_j}}}\,,\,\,\,\,\,\,\,\,\,\,\,i,\,j = 1,2
	\label{momentum}
\end{equation}

\noindent and continuity
\begin{equation}
	\frac{{\partial \,{v_i}}}{{\partial \,{x_i}}} = 0,\,\,\,\,\,\,\,\,\,\,\,\,\,i = 1,2.
	\label{continuity}
\end{equation}

As \textit{initial conditions}, zero values are set for all the dependent variables. \textit{Boundary conditions} consist in no-slip constraints at the walls \begin{math} {u_{wall}  = v_{wall}  = 0} \end{math},  setting the outlet pressure to 0, and increase of the inlet pressure from 0 up to 1 on  a smooth law  to reach a quasi-steady flow (cf. \cite{Mackarov_arxiv} ):
\begin{equation}
	{p_{inlet}(t)={\alpha t} \mathord{\left/
			{\vphantom {{ kt} {(1 +\alpha t)}}} \right.
			\kern-\nulldelimiterspace} {(1 + \alpha t)}}.
\end{equation}
The results below are all gotten with $\alpha=1$ (cf. \cite{Mackarov2011}). When $t =25$, then the inlet pressure time derivative is less than 0.15\% of its initial value. So later on, the flow conditions are considered near-stationary.

\section{Asymptotic solution for the symmetric stationary spread over a wall}
\label{sec:asymp:spread}

Study now a flow moving along a horizontal slot to meet a vertical wall and  spread over it (Fig.~\ref{fig:spreadLayout}).

Similarly to the counterflows case \cite{Mackarov_arxiv,Mackarov2011}, we can construct a solution for the UCM fluid spread over a wall near the stagnation point $x=y=0$.

Besides the no-slip condition
\begin{equation}
	u(0,y) = v(0,y)=0,
	\label{spread:slip}
\end{equation}
the flow with the symmetry corresponding to the domain symmetry (relative to the \textit{x} axis) should evidently satisfy following conditions for the velocities and pressure:
\begin{gather}
	u(x,y) = u(x, - y),
	\label{spread:u} \\
	v(x,y) =  - v(x, - y),
	\label{spread:v}\\
	p(x,y) = p(x, - y).
	\label{spread:p}
\end{gather}
Next, since $\sigma _{xx}$ and $\sigma _{xy}$ are the fluxes of the \textit{x} and \textit{y} components of the mechanical momentum across a unit area normal to the  \textit{x} axis, they must obey the  conditions of symmetry and antisymmetry, respectively, relative to \textit{x}:
\begin{gather}
	{\sigma _{xx}}(x,y) = {\sigma _{xx}}(x, - y),
	\label{spread:sxx}\\
	{\sigma _{xy}}(x,y) =  - {\sigma _{xy}}(x, - y).
	\label{spread:sxy}
\end{gather}
Likewise,
\begin{equation}
	{\sigma _{yy}}(x,y) = {\sigma _{yy}}(x, - y).
	\label{spread:syy}
\end{equation}

\indent Write out the most general asymptotic power representation of solutions to the velocities near the central point satisfying the continuity equation (\ref{continuity}) and conditions (\ref{spread:slip})--(\ref{spread:v}). The one up to the terms of the third order with respect to  \textit{x} and \textit{y } looks like
\begin{gather}
	u\left( {x,y} \right) = A  x^2 + B  x^3,
	\label{spread:u.symmetric}\\
	v\left( {x,y} \right) =  - 2 A x y - 3 B {x^2}y.
	\label{spread:v.symmetric}
\end{gather}
From conditions (\ref{spread:sxx}) and (\ref{spread:syy}) the general stationary third-order form follows of the normal stresses expansions:
\begin{gather}
	\sigma_{xx} \left( {x,y} \right) = {\Sigma}{_x} + {\alpha}x + {\rm{\beta}}{x^2} + {\rm{\delta}}{y^2} + {\rm{\varepsilon}}x{y^2} + \zeta {x^3},
	\label{spread:sigma.xx}\\
	\sigma_{yy} \left( {x,y} \right) = {\Sigma}{_y} + {\eta}x + {\rm{\theta}}{x^2} + {\rm{\iota}}{y^2} + {\rm{\kappa}}x{y^2} + \lambda {x^3}.
	\label{spread:sigma.yy}
\end{gather}
\noindent
Condition (\ref{spread:sxy})  gives a representation of the shear stress:
\begin{equation}
	{\sigma _{xy}}\left( {x,y} \right){\rm{ = }}\mu y + \xi xy + \varrho {x^2}y + \tau {y^3}.
	\label{spread:sigma.x}
\end{equation}
\noindent
Lastly, compose an expression for the pressure taking account of the momentum equation (\ref{momentum}) and condition (\ref{spread:p}):
\begin{equation}
	p\left( {x,y} \right) = {P_0} + {P_x}x + {P_{xx}}{x^2} + {P_{yy}}{y^2} + {P_{xxx}}{x^3} + {P_{xyy}}x{y^2}.
	\label{spread:p.symmetric}
\end{equation}
Similarly to \cite{Mackarov_arxiv}, we obtain these expansions assuming existence of all the variables' fourth derivatives.

Substitute relations (\ref{spread:u.symmetric})-(\ref{spread:p.symmetric}) into system (\ref{UCM})-(\ref{continuity}) and  compare the same order terms. This
\begin{itemize}[label=\textbullet]
	\item brings about a condition $A=0$,
	\item makes determine their coefficients and obtain final expressions for the flow variables containing \emph{B} as a known parameter (similarly to the asymptotic solution to the counterflows, it must be determined by merging the asymptotic and global solutions to the spread problem):
	\begin{gather}
		u\left( {x,y} \right) = B  x^3,
		\label{Spread:u:final}\\
		v\left( {x,y} \right) = - 3 B {x^2}y,
		\label{Spread:v:final}\\
		\sigma_{xx} \left( {x,y} \right) = \frac{{6B{x^2}}}{\Rey},
		\label{Spread:sxx:final}\\
		\sigma_{yy} \left( {x,y} \right) = - \frac{{6B{x^2}}}{\Rey},
		\label{Spread:syy:final}\\
		\sigma_{xy} \left( {x,y} \right) = - \frac{{6B{xy}}}{\Rey},
		\label{Spread:sxy:final}\\
		p \left( {x,y} \right) = {P_0} + \frac{{3B~{\left(x^2-y^2\right)}}}{\Rey}.
		\label{Spread:p:final}
	\end{gather}
\end{itemize}

Mention now the statement of \cite{Becherer,Objections} in regard with the elongational and wall stagnation flows: these flows' stationary points definitely correspond to singularities in the solutions to the UCM constitutive state equation. The asymptotic analytical solution of ours does not confirm this statement \emph{with respect to the wall stagnation flow}.  As follows from Eqs.~(\ref{Spread:u:final})-(\ref{Spread:p:final}), the symmetric solution is regular, unique, and strictly satisfies the problem governing equations (including the UCM state equation) together with the boundary/symmetry conditions.
More evidence of the solution regularity is given below.

\section{Numerical analysis of the UCM fluid spread}
\label{sec:NumSpread}
\begin{figure*}[t]
	\centering
	\begin{subfigure}[b]{0.1485\textwidth}
		\includegraphics[width=1\textwidth]{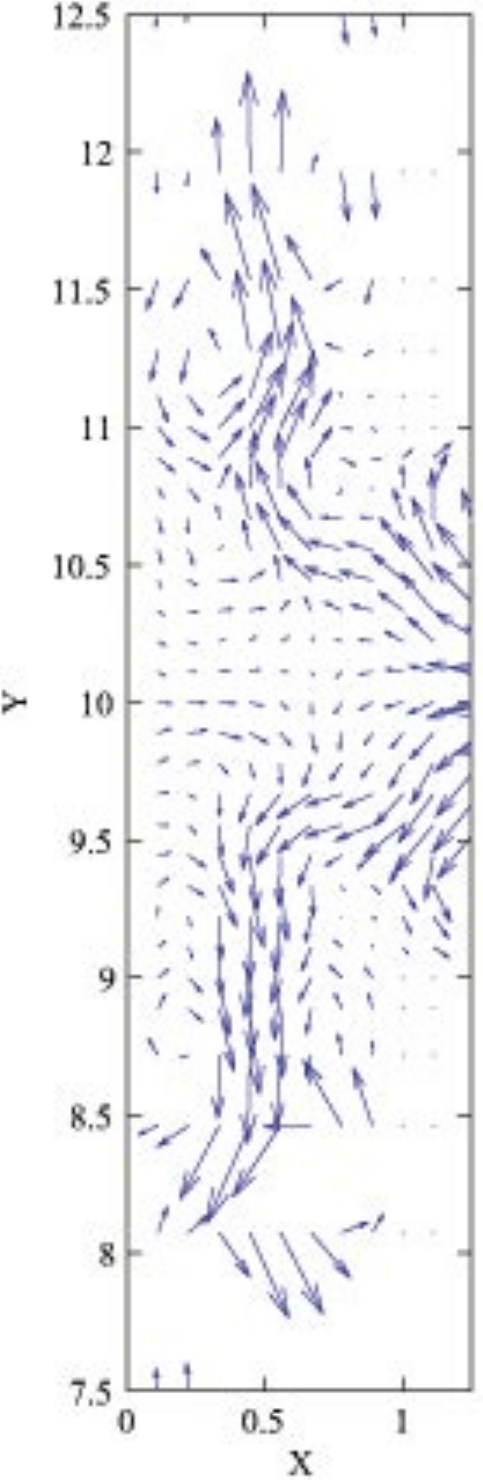}
		\caption{\scriptsize{$\Delta t = 5\cdot10^{-4}$,\\ $ \Delta = 0.10$.}}
	\end{subfigure} ~~
	\begin{subfigure}[b]{0.1515\textwidth}
		\includegraphics[width=1\textwidth]{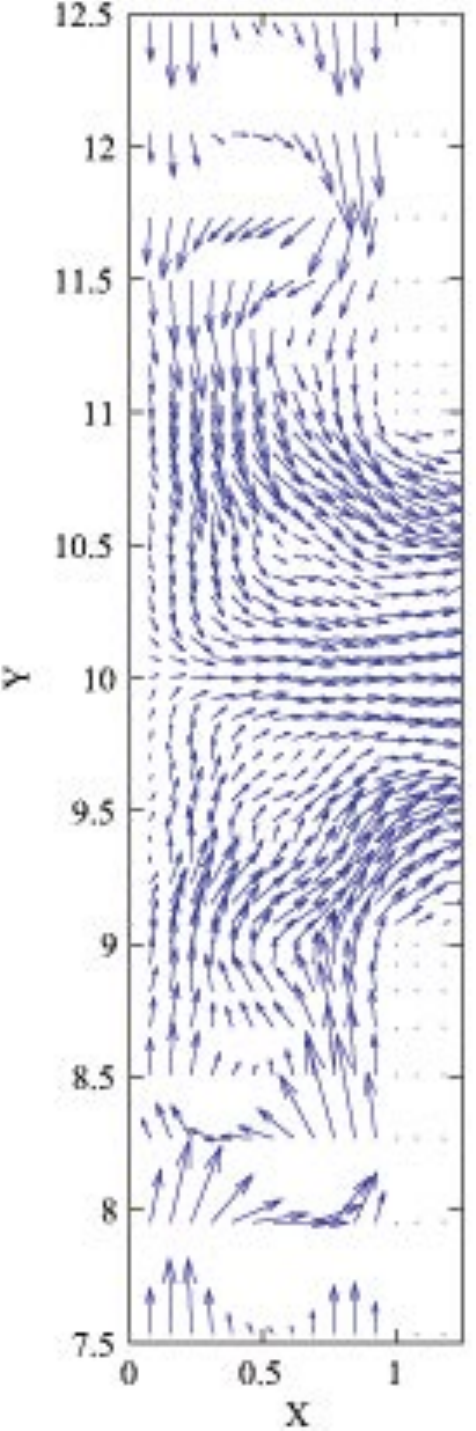}
		\caption{\scriptsize{$\Delta t =5\cdot{10^{-4}}$,\\  $\Delta = 0.071$}}
	\end{subfigure} ~~
	\begin{subfigure}[b]{0.1505\textwidth}
		\includegraphics[width=1\textwidth]{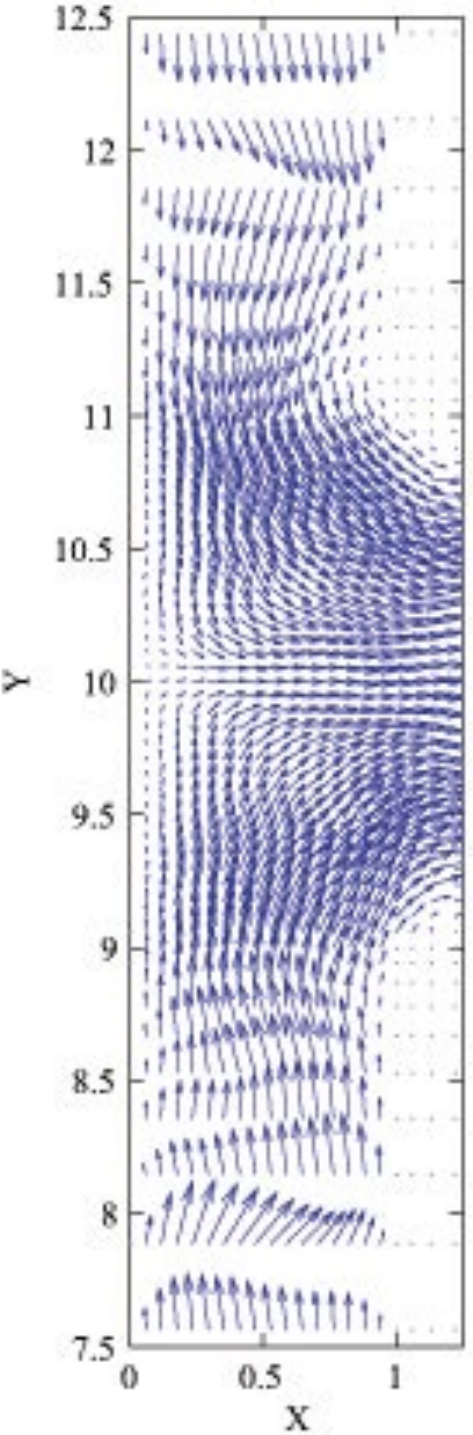}
		\caption{\scriptsize{$\Delta t =10^{-4}$,\\  $\Delta = 0.056$.}}
	\end{subfigure} ~~
	\begin{subfigure}[b]{0.1495\textwidth}
		\includegraphics[width=1\textwidth]{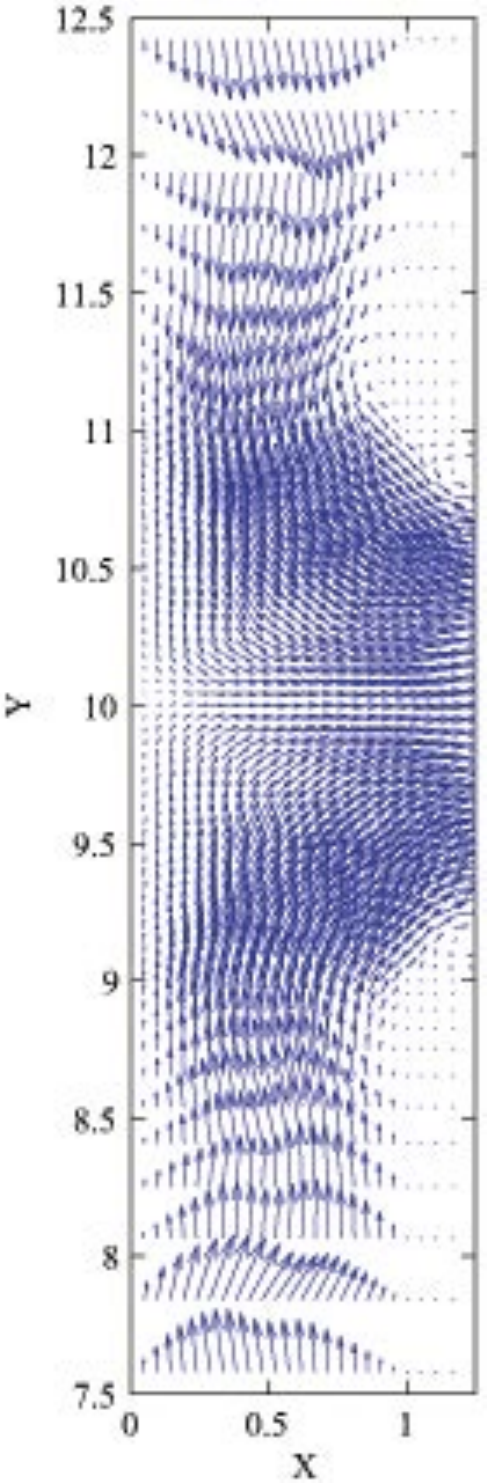}
		\caption{\scriptsize{$\Delta t =10^{-4}$, \\  $\Delta = 0.046$}}
	\end{subfigure}
	\caption{An instance of the spread over a wall  given by the numerical simulation on different meshes; $t=4.045$, $\Rey=0.003$, $\Wi=150$. Specified are the time and mesh steps, $\Delta t$ and $ \Delta$.}
	\label{fig:VortexSPreadPattern}
\end{figure*}

\begin{figure*}[t]
	\centering
	\begin{subfigure}[b]{0.1495\textwidth}
		\includegraphics[width=1\textwidth]{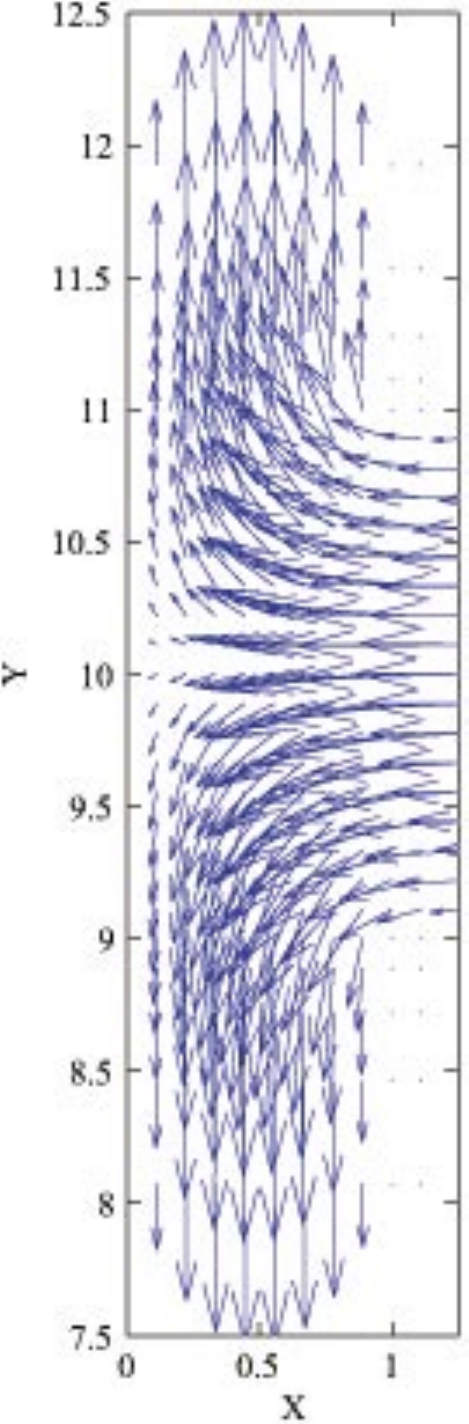}
		\caption{\scriptsize{$\Delta t = 10^{-3}$,\\ $ \Delta = 0.10$}}
	\end{subfigure} ~~
	\begin{subfigure}[b]{0.1499\textwidth}
		\includegraphics[width=1\textwidth]{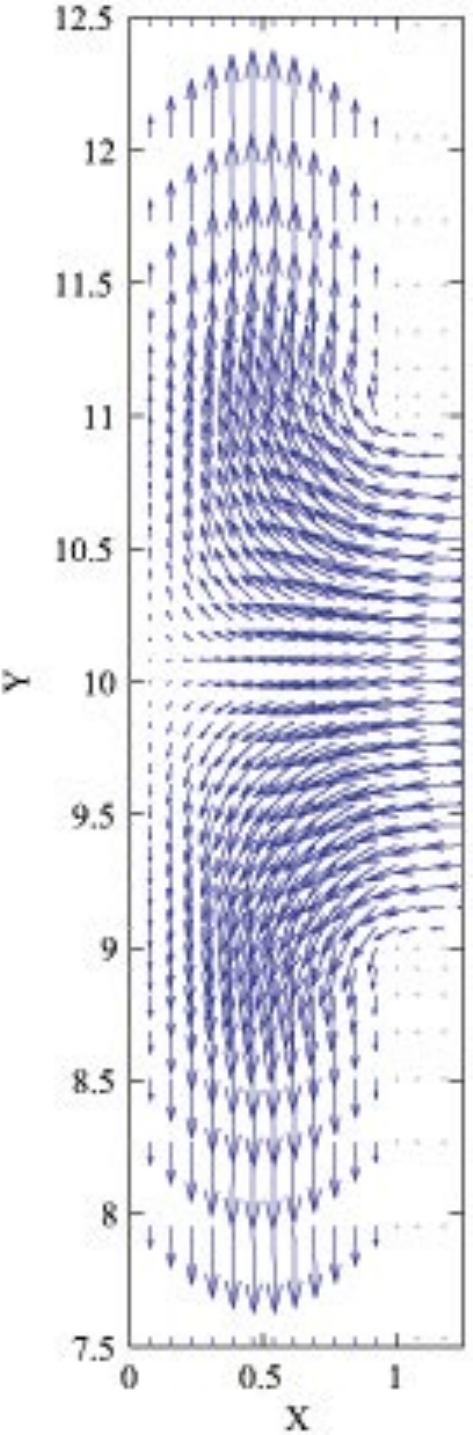}
		\caption{\scriptsize{$\Delta t =5\cdot{10^{-4}}$,\\  $\Delta = 0.071$}}
	\end{subfigure} ~~
	\begin{subfigure}[b]{0.1503\textwidth}
		\includegraphics[width=1\textwidth]{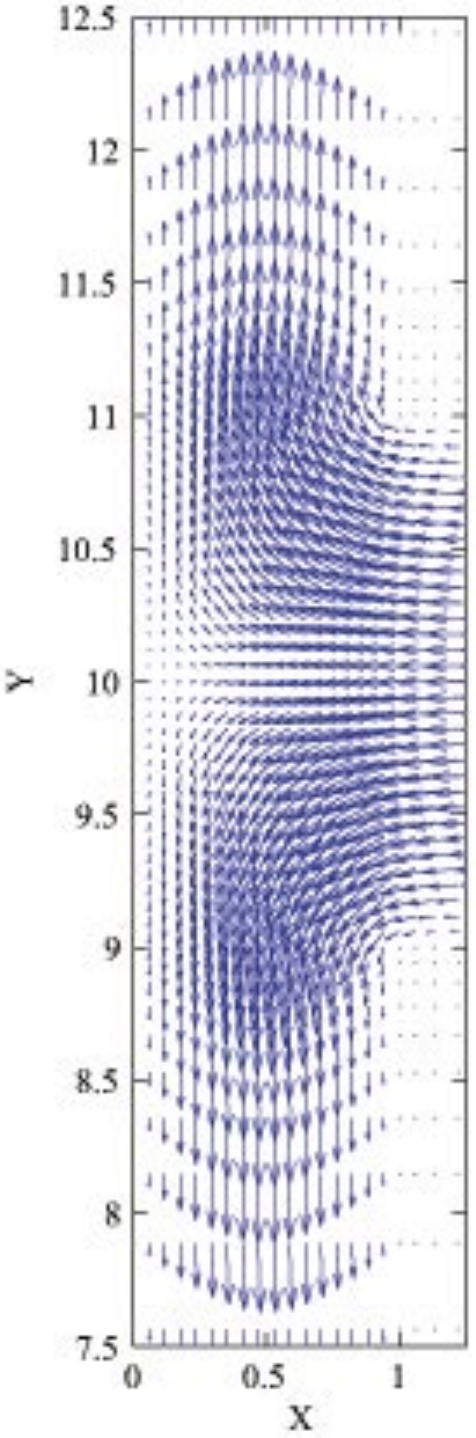}
		\caption{\scriptsize{$\Delta t =10^{-4}$,\\  $\Delta = 0.056$}}
	\end{subfigure} ~~
	\begin{subfigure}[b]{0.1503\textwidth}
		\includegraphics[width=1\textwidth]{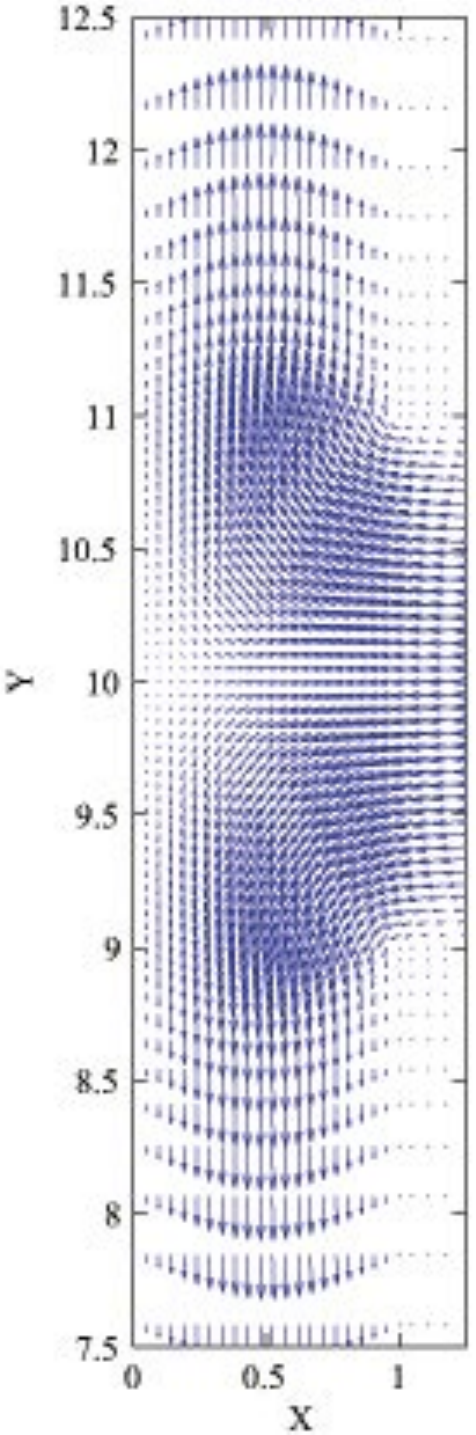}
		\caption{\scriptsize{$\Delta t =10^{-4}$, \\  $\Delta = 0.046$}}
	\end{subfigure}
	\caption{An instance of the spread over a wall  given by the numerical simulation on different meshes; $t=5.85$, $\Rey=0.003$, $\Wi=150$. Specified are the time and mesh steps, $\Delta t$ and $ \Delta$.}
	\label{fig:RegularSPreadPattern}
\end{figure*}

\vspace*{0.2cm}
Like the counterflows, the spread over a wall on its transient phase is accompanied  by periodic formation of vortex-like structures and change of the flow direction (as in Fig.~\ref{fig:VortexSPreadPattern}), with the flow getting regular in between (Fig.~\ref{fig:RegularSPreadPattern}). This is not surprising since a good deal of this flow moves within the straight channels too, and the acceleration phase of a flow inside such channels \textit{must} be accompanied by reversals as has been shown in \cite{Mackarov2011}.

\begin{figure*}[t]
	\centering
	\includegraphics[width=0.7\textwidth]{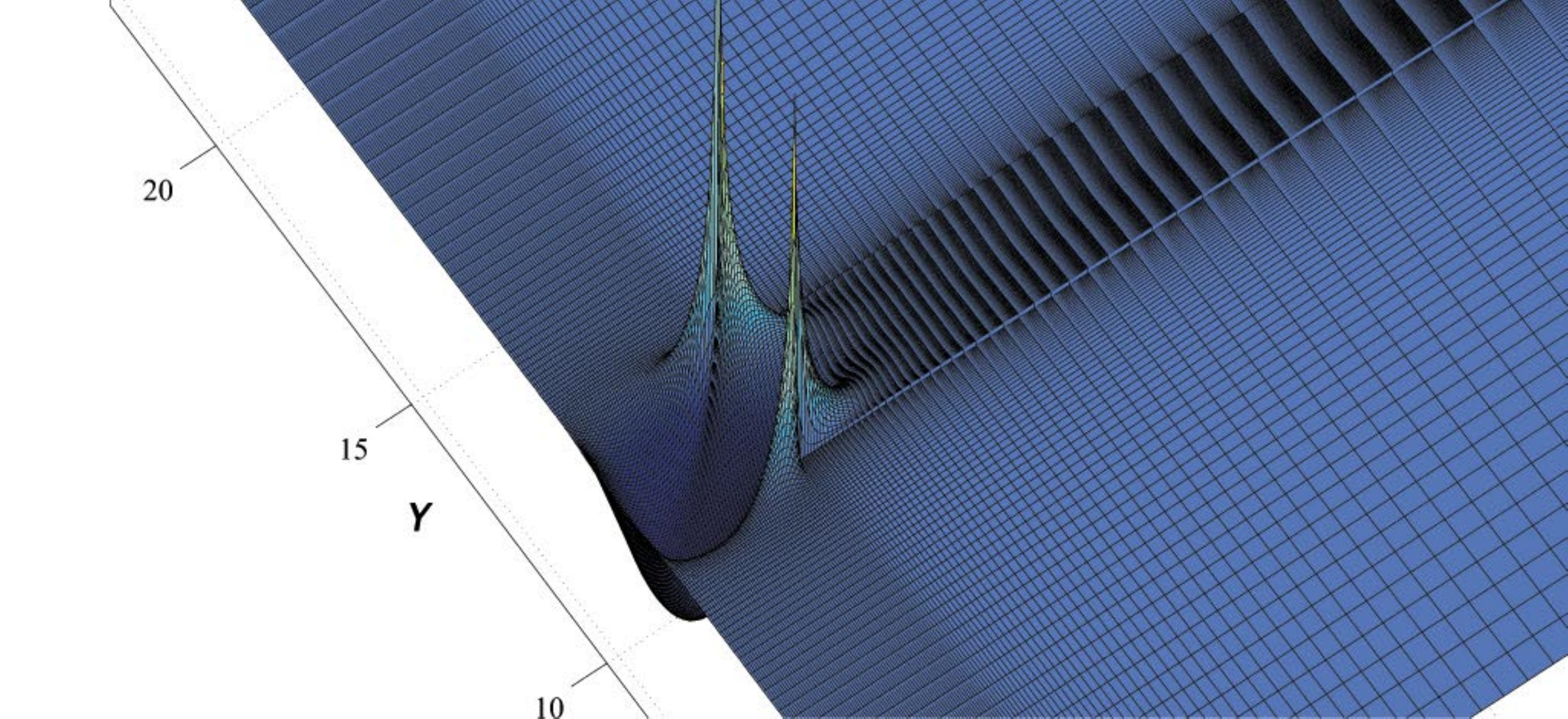}
	\caption{Profile of the  spreading flow normal stress $\sigma_{xx}$ for $ \Rey = \Wi = 1$ on the stationary regime ($t=30$). The mesh minimum step is 0.017, $\Delta t=10^{-4}$.}\label{fig:Stress-60-60}
\end{figure*}

At the same time, the spread over a wall proved to get restructured  more rapidly and intensively than the counterflows. Overall, the spread flow is characterized by a ``sharper'' distribution of its characteristics compared to the  counterflows. Such is, for instance, a pretty distinct minimum of the normal stress seen in Fig.~\ref{fig:Stress-60-60}. It is usually located \textit{circa} midway between the stagnation point and walls corners. With high Weissenberg numbers, it may be split to form a couple of local minima.

Such a complicated structure, in particular, required much more iterations for the pressure correction method to reach the incompressible fluid state (see subsection \ref{PCM}) and a smaller time step. We will still discuss this flow peculiarities in detail in Section \ref{issues}.

Regarding Fig.~\ref{fig:Stress-60-60}~note that, unlike the non-zero value of the normal stress in the center of the counterflows, the spreading flow normal stress tends to 0 (not infinity!) with small $x,\ y$ just as suggested by Eq. (\ref{Spread:sxx:final})

\subsection{Convergence of the numerical solution}

\begin{table*} 
	\centering
	\caption{Sequence of normal stress minimum values for the spread over a wall \textit{vs.} a total number of mesh nodes (or the least space step for each mesh) and a time step; $t = 30$, $\Rey=1$,\ $\Wi=1$. }
	\label{tab:1}
	\scriptsize{\begin{tabular}{|l|||l|l|l|l|l|l|l|l|l|}
		\hline
		\multicolumn{1}{|c||}{\emph{N}}  & \multicolumn{1}{c|}{283} & \multicolumn{1}{c|}{613} & \multicolumn{1}{c|}{751} & \multicolumn{1}{c|}{903} & \multicolumn{1}{c|}{1443} & \multicolumn{1}{c|}{ 1651 } & \multicolumn{1}{c|}{1873} & \multicolumn{1}{c|}{6033} & \multicolumn{1}{c|}{24663} \\
		\hline
		\multicolumn{1}{|c||}{$h_{min}$}  & \multicolumn{1}{c|}{0.14} & \multicolumn{1}{c|}{0.1} & \multicolumn{1}{c|}{0.091} & \multicolumn{1}{c|}{0.083} & \multicolumn{1}{c|}{0.071} & \multicolumn{1}{c|}{0.067} & \multicolumn{1}{c|}{0.059} & \multicolumn{1}{c|}{0.01} & \multicolumn{1}{c|}{0.017} \\
		\hline
		\multicolumn{1}{|c||}{$\Delta t$ } & \multicolumn{1}{c|}{0.00005} & \multicolumn{1}{c|}{0.00005} & \multicolumn{1}{c|}{0.00005} & \multicolumn{1}{c|}{0.00005} & \multicolumn{1}{c|}{0.00003} & \multicolumn{1}{c|}{0.00003} & \multicolumn{1}{c|}{0.000025} & \multicolumn{1}{c|}{0.00001} & \multicolumn{1}{c|}{0.000001} \\
		\hline
		\multicolumn{1}{|c||}{$\sigma_{xx\ min}$} &  \multicolumn{1}{c|}{-0.0224} & \multicolumn{1}{c|}{-0.0215} & \multicolumn{1}{c|}{-0.0219} & \multicolumn{1}{c|}{-0.0219} & \multicolumn{1}{c|}{-0.0218} & \multicolumn{1}{c|}{-0.0217} & \multicolumn{1}{c|}{-0.0216} & \multicolumn{1}{c|}{-0.023} & \multicolumn{1}{c|}{-0.024} \\
		\hline
	\end{tabular}}
\end{table*}

Following \cite{Mackarov_arxiv,Mackarov2011}, analyze the numerical solution convergence looking at one of its ``special'' points, where the solution rapidly changes in time and space. This is a  point of a distinct normal stress $\sigma_{xx}$ absolute minimum  referred in the previous section and observed in all the experiments.
\begin{table*}
	\centering
	\caption{Sequence of normal stress absolute minima for the spread over a wall versus a total number of mesh nodes (or the least space step for each mesh) and a time step; $t = 30$, $\Rey=0.05$, $\Wi=4$. }
	\label{tab:2}
	\scriptsize{
	\begin{tabular}{|l||l|l|l|l|l|l|l|l|l|}
		\hline
		\multicolumn{1}{|c||}{\emph{N}} & \multicolumn{1}{c|}{133} & \multicolumn{1}{c|}{379} & \multicolumn{1}{c|}{613} & \multicolumn{1}{c|}{751} & \multicolumn{1}{c|}{903} & \multicolumn{1}{c|}{1249} & \multicolumn{1}{c|}{1443} & \multicolumn{1}{c|}{1873} & \multicolumn{1}{c|}{6033} \\
		\hline
		\multicolumn{1}{|c||}{$h_{min}$} & \multicolumn{1}{c|}{0.2} & \multicolumn{1}{c|}{0.125} & \multicolumn{1}{c|}{0.1} & \multicolumn{1}{c|}{0.091} & \multicolumn{1}{c|}{0.083} & \multicolumn{1}{c|}{0.071} & \multicolumn{1}{c|}{0.067} & \multicolumn{1}{c|}{0.059} & \multicolumn{1}{c|}{0.01} \\
		\hline
		\multicolumn{1}{|c||}{$\Delta t$ } & \multicolumn{1}{c|}{0.00001} & \multicolumn{1}{c|}{0.00001} & \multicolumn{1}{c|}{0.00001} & \multicolumn{1}{c|}{0.00001} & \multicolumn{1}{c|}{0.00001} & \multicolumn{1}{c|}{0.00001} & \multicolumn{1}{c|}{0.00001} & \multicolumn{1}{c|}{0.00001} & \multicolumn{1}{c|}{0.00001} \\
		\hline
		\multicolumn{1}{|c||}{$\sigma_{xx\ min}$} & \multicolumn{1}{c|}{-0.0231} & \multicolumn{1}{c|}{-0.0199} & \multicolumn{1}{c|}{-0.0193} & \multicolumn{1}{c|}{-0.0197} & \multicolumn{1}{c|}{-0.0198} & \multicolumn{1}{c|}{-0.0198} & \multicolumn{1}{c|}{-0.0196} & \multicolumn{1}{c|}{-0.0196} & \multicolumn{1}{c|}{-0.0196} \\
		\hline
	\end{tabular}}
\end{table*}

The case of a fluid with a moderate or small viscosity, with essential\textit{ inertial instability} factor, is presented by Table \ref{tab:1}. This case turned out rather hard to calculate. Just as in the case of the counterflows with $\Wi=\Rey=1$ \cite{Mackarov_arxiv,Mackarov2011}. In particular, each concrete mesh required a thorough choice of a time step to provide convergence. We see, however, a rather regular sequence of the stress values, probably except for two most finest  meshes.  

This conceivably corresponds to a known phenomenon of a noticeable accumulation of computational errors in fine mesh solutions to evolutionary problems of fluid dynamics, which was discussed in \cite{Mackarov2000,Mackarov2011}. With respect to the spread over a wall, given that the flow field has an especially fine-grained structure, a very fine mesh may reveal the flow's subtle details, averaged and fuzzy on rougher meshes. This can make the finer mesh stress values deflect from the general trend, to some extend.

When both elasticity and viscosity are moderate (Table \ref{tab:2}), the stress values form quite a regular sequence, except for especially rough meshes.

Lastly, Table \ref{tab:3} demonstrates the spread flow with \textit{a high Weissenberg number} as essentially dynamic and non-stationary. Like the highly elastic fluid counterflows, it is, in particular, sensitive to both the time and mesh steps. Using a bigger time step or a finer mesh (the italicized columns)  puts the minimum stress out of its main trend formed up by the rest of the columns. Such a violation is not so critical though. The solutions are still regular and contain main principal features of spread flows in general as seen in Figs.~\ref{fig:VortexSPreadPattern}, \ref{fig:RegularSPreadPattern}.

\subsection{Details of the numerical procedure}
\label{sec:NumDetails}

\subsubsection{Mesh}
Both the counterflows and spread layouts (Fig.~\ref{fig:layout}) have a quadrangle located between the other parts of the computational domain, horizontal and vertical slots (four or three, respectively).

All the five or four domain parts  have equal numbers of the mesh steps on \textit{x} and \textit{y}. The mesh steps are hereby uniform within the central quadrangle, and  flexibly (in geometric progression) diminish towards the quadrangle (and towards the value of a step inside it).

\begin{table*}
	\centering
	\caption{Sequence of normal stress absolute minima for the spread over a wall versus a total number of mesh nodes (or the least space step for each mesh) and a time step; $t = 30$, $\Rey=0.03$, $\Wi=150$. }
	\label{tab:3}
	\begin{tabular}{|l||l|l|l|l|l|l|l|}
		\hline
		\multicolumn{1}{|c||}{\emph{N}} & \multicolumn{1}{c|}{613} & \multicolumn{1}{c|}{751} & \multicolumn{1}{c|}{903} & \multicolumn{1}{c|}{\textit{1443}} & \multicolumn{1}{c|}{ 1651 } & \multicolumn{1}{c|}{1873} & \multicolumn{1}{c|}{\textit{6033}} \\
		\hline
		\multicolumn{1}{|c||}{$h_{min}$} & \multicolumn{1}{c|}{0.1} & \multicolumn{1}{c|}{0.091} & \multicolumn{1}{c|}{0.083} & \multicolumn{1}{c|}{\textit{0.071}} & \multicolumn{1}{c|}{0.067} & \multicolumn{1}{c|}{0.059} & \multicolumn{1}{c|}{\textit{0.01}} \\
		\hline
		\multicolumn{1}{|c||}{$\Delta t$} & \multicolumn{1}{c|}{0.00001} & \multicolumn{1}{c|}{0.00001} & \multicolumn{1}{c|}{0.00001} & \multicolumn{1}{c|}{\textit{0.00003}} & \multicolumn{1}{c|}{0.00001} & \multicolumn{1}{c|}{0.00001} & \multicolumn{1}{c|}{\textit{0.00001}} \\
		\hline
		\multicolumn{1}{|c||}{$\sigma_{xx\ min}$} & \multicolumn{1}{c|}{-0.0104} & \multicolumn{1}{c|}{-0.0108} & \multicolumn{1}{c|}{-0.0104} & \multicolumn{1}{c|}{\textit{-0.0160}} & \multicolumn{1}{c|}{-0.0107} & \multicolumn{1}{c|}{-0.0107} & \multicolumn{1}{c|}{\textit{-0.023}} \\
		\hline
	\end{tabular}
\end{table*}

\begin{figure*}[t]
	\centering
	\begin{subfigure}[b]{0.3015\textwidth}
		\includegraphics[width=1\textwidth]{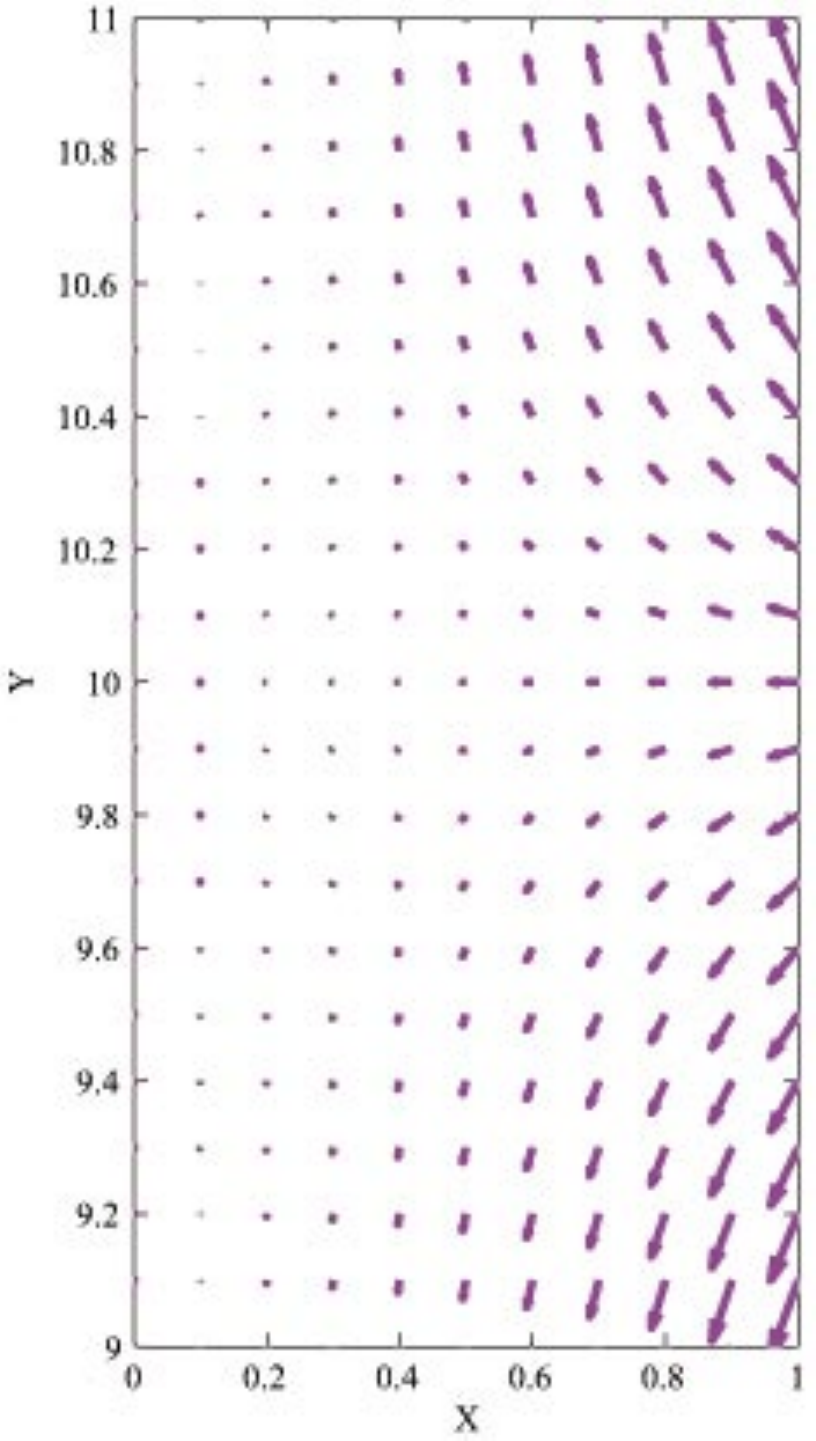}
		\caption{ \scriptsize{the flow field given by Eqs.~ (\ref{Spread:u:final}),\ (\ref{Spread:v:final}) with $B=-1$}}
		\label{fig:Comparison:analyt}
	\end{subfigure} ~~
	\begin{subfigure}[b]{0.2978\textwidth}
		\includegraphics[width=1\textwidth]{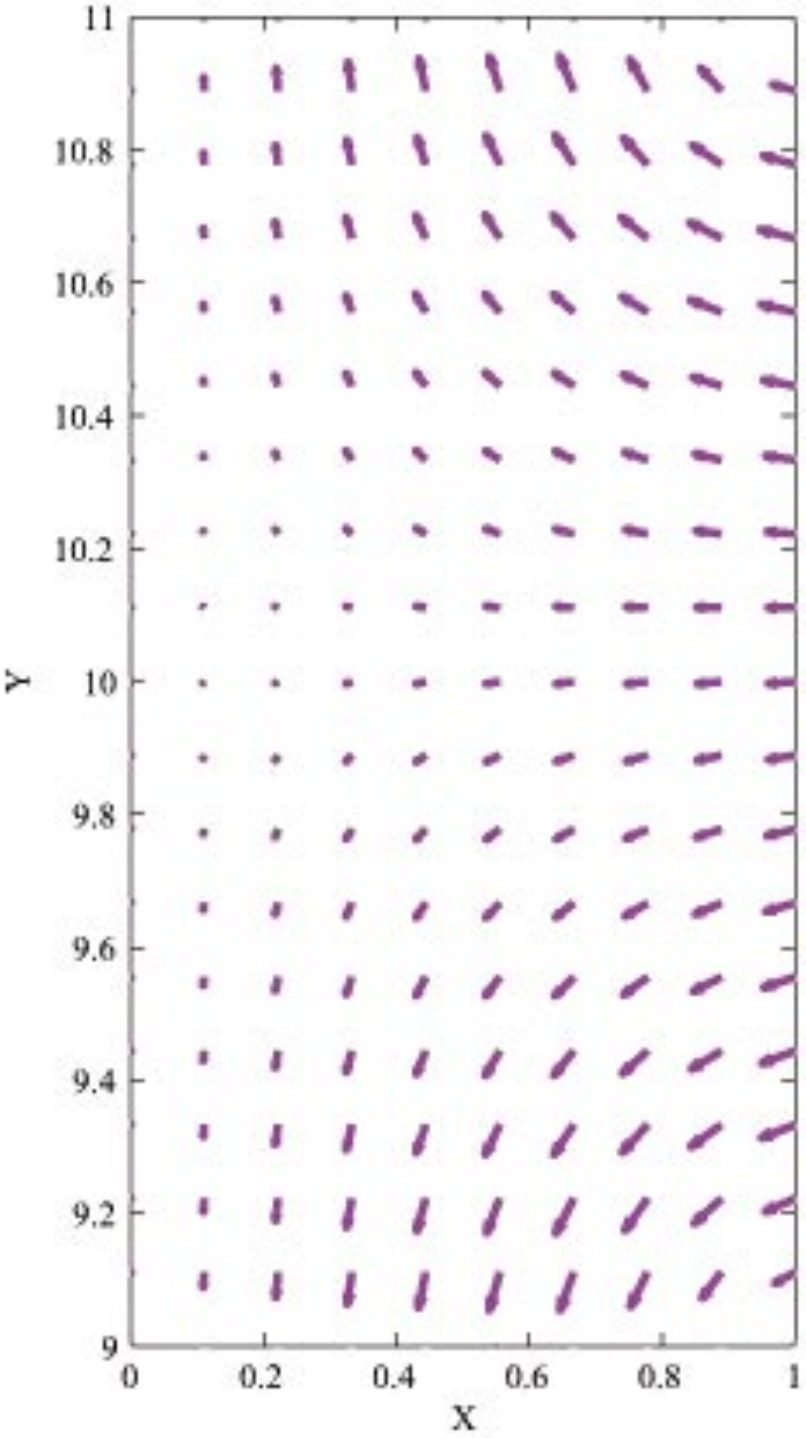}
		\caption{\scriptsize{the numerical flow field \\ ($\Rey=1,\ \Wi=1,\ t=30$)}}
	\end{subfigure}
	\caption{Comparison of the analytical and numerical spread flow patterns. }
	\label{fig:Comparison}
\end{figure*}

\subsubsection{Computational scheme}
In the evolutionary equations an explicit first-order finite-difference approximation of the time derivatives was used. In the momentum equation the pressure gradient was approximated by the backward finite differences. In the UCM state equation the spatial derivatives of the stresses and velocities were described by the forward finite differences, while for the derivatives of the stresses in the momentum equation  the backward differences were again applied.  The convective terms in the momentum equation were represented by the upwind finite differences.

Such a choice of finite differences was dictated, \textit{inter alia}, by usage of the pressure correction method.

\subsubsection{Pressure correction method}
\label{PCM}
The applied numerical procedure is essentially based upon a simple and pretty efficient
version of the pressure correction method (hereafter also PCM). With its description and convergence proof available in \cite{Mackarov_arxiv}, discuss here its features and overall value that concern the present work.

At each time step the iterative procedure of PCM adjusts the pressure and velocity so as to satisfy the continuity equation and thus to attain the incompressible fluid state. The iterations were terminated when the norm \cite{Mackarov_arxiv} of the velocity divergence reached the value of $10^{-5}$, which is much smaller than the smallest step on a space variable for any of the meshes used in practice. Thus the error invaded by PCM was negligent compared to the space finite approximation errors.

Herewith in the case of the counterflows it usually took PCM 1--2 iterations to correct the pressure on stable flow phases, whereas on top of the vortexes' intensities 10--12 ones were often needed.

As to the spread flow, much more iterations were sometimes required: up to 200-300 for $t<1$, with the least stationary flow. A slower convergence conceivably follows from a specific, more fine-grained structure of this kind of a flow, which will yet be analyzed below.

In relation with the hitherto stated opinions about singularities as a characteristic feature of the UCM model \textit{per se} or poor compliance between the UCM rheological law and momentum equation, a word must be said about a possible role of PCM in obtaining regular solutions in this research. At any PCM iteration the corrected pressure and velocity values in every node depend upon the pressure and velocity in its neighboring nodes, which in their turn are affected by the joint nodes values, etc.  Thus, one can say about \textit{integral}, or \textit{hyperbolic}  nature of PCM. It seems to be able to ``smear out'' sharp short-wave disturbances (once they evidently can  corrupt the incompressibility \textit{locally}) --- specific for high elasticity --- over the whole of the flow domain. It is this feature of PCM that most probably makes the method successful in dealing with high Weissenberg numbers.

No doubt, this PCM trait is worth deeper and more formal investigation in the future.

\subsection{On some issues of the numerical solution near the stagnation point}
\label{issues}

Regard the remark of the previous section about more fine and sophisticated distribution of the spread flow field compared to the counterflows. Eqs. (\ref{Spread:u:final}), (\ref{Spread:v:final}) indicate that the velocities are cubic on $x,\ y$ near the stationary point, unlike the case of the counterflows, whose velocities  can generally have the first order on the space variables  near the center. Consequently, in terms of the spread flow, a first-order or even a second-order velocity field approximation may evidently result in an error comparable with the leading part. So, a numerical simulation like this may not guarantee resolution of every finest detail of the solution in the vicinity of the stationary point. Evidently this is only pertinent to the region of small velocities near the stationary point and cannot result in a false indication of regularity instead of singularity or \emph{vice versa}.
\begin{figure*}[t]
	\centering
	\includegraphics[width=1\textwidth]{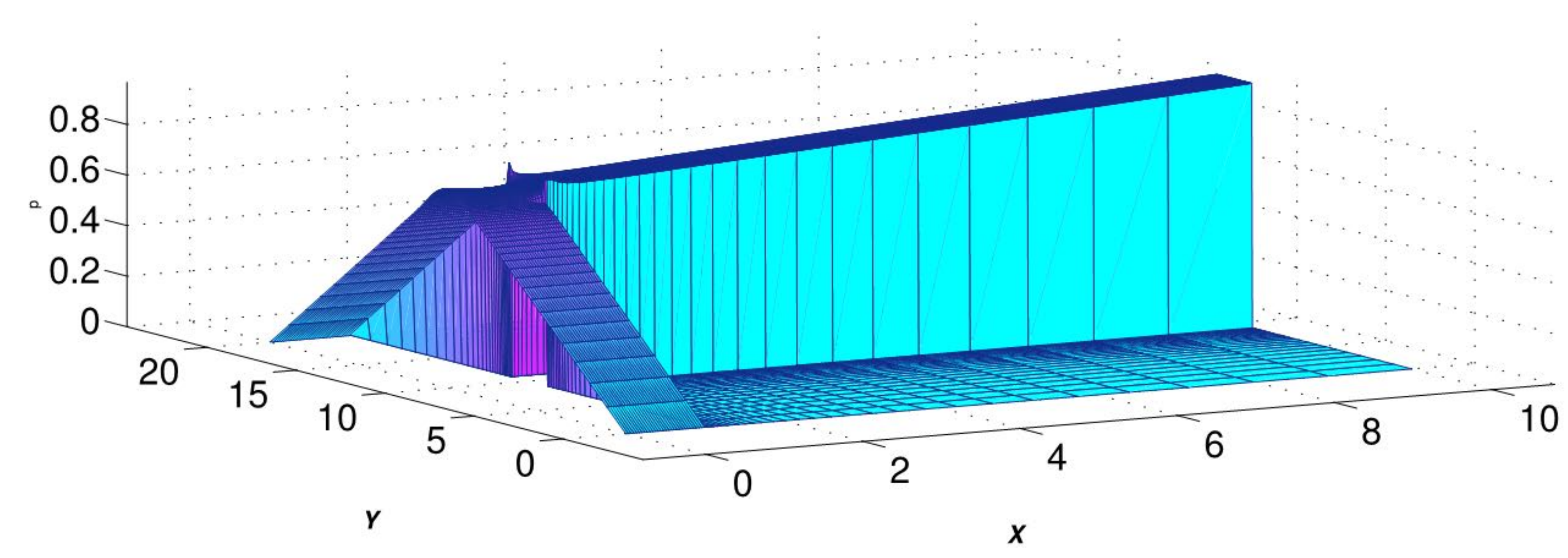}
	\caption{Profile of the spreading flow pressure at $ \Rey = \Wi = 1$ on the stationary regime ($t=29.25$). The mesh minimum step is 0.034, $\Delta t=10^{-4}$.}\label{fig:Pressure-30-30}
\end{figure*}

Such a region is shown in Fig.~\ref{fig:Comparison}. For the analytical flow pattern \ref{fig:Comparison:analyt}~ its only parameter \textit{B} is chosen such that both patterns look similar closer to the borders of this region. At the same time, one can notice that on the way to the stagnation point the analytical velocities tend to zero a little faster. Obviously the numerical solution tries to alleviate this tendency.

Lastly, mention another point of the numerical solutions unusual behavior. This is a corner between the walls. In both problems considered very high pressure and stress gradients are usually there observed (especially high with smaller Reynolds numbers). Interestingly, the pressure  and stress gradients nearly balance each other so that the velocity field is smooth and satisfies the no-slip conditions (Figs.~\ref{fig:Stress-60-60}, \ref{fig:Pressure-30-30}). A thorough look at the solutions reveals that such peculiarities of the solution at the corners are mostly local and do not remarkably affect the flow as a whole and the situation near the stagnation point in particular.

\section{Conclusions and discussion}

On the ground of the presented research we thus can confirm the paper's title statement.

Indeed, two typical benchmark flows of the UCM fluid were investigated using an analytical and numerical procedures.  Both approaches were based on rigorous meeting the governing relations,\textit{ viz}. the momentum, continuity, and UCM constitutive state equations, as well as physically natural boundary conditions. As far as the momentum equation is concerned, its convective terms were always kept, since even for high viscosities they were not negligible at least near stagnation points and inside vortexes because of high strains.

So there was no need to overburden the problems statement by additional pre-study suppositions (for example,  about a flow velocity field or distributions of the stresses over the space \cite{VanGorder}). As a result, clearly regular flows were observed in the region of stationary points.

Thus, singular stresses are not believed to be an intrinsic feature of the UCM rheological model, which does not mean, of course, that it never brings about sharp peaks of stresses. Such are the distinct stresses extrema detected in both kinds of the flows at the walls corners.

Though this special  behavior of the  stresses does not seem to affect regularity of a flow near a stagnation point, a flow of a viscoelastic fluid near a corner as such is an interesting subject matter for future investigations. Especially attractive would be finding an analytical solution to the flow in a region like this. No doubt, such a research would generally deepen the comprehension of viscoelasticity.

\bibliographystyle{plain}
\bibliography{Mackarov}
\end{document}